\documentclass[aps,prb,twocolumn,showpacs]{revtex4}
\usepackage{epsfig,graphicx}
\usepackage{amssymb}
\begin{document}
\title{A fast and reliable method for the calculation of band structure of
solids with hybrid functionals}
\author{Fabien Tran}
\affiliation{Institute of Materials Chemistry, Vienna University of Technology,
Getreidemarkt 9/165-TC, A-1060 Vienna, Austria}

\begin{abstract}

A simple approximation within the framework of the hybrid methods for the
calculation of the electronic structure of solids is presented. By considering
only the diagonal elements of the perturbation operator (Hartree-Fock exchange
minus semilocal exchange) calculated in the basis of the semilocal orbitals,
the computational time is drastically reduced, while keeping very well in most
studied cases the accuracy of the results obtained with hybrid functionals when
applied without any approximations.

\end{abstract}

\maketitle

Quantum calculations for solids are commonly done with the local density
approximation (LDA) or the generalized gradient approximation (GGA) of density-functional
theory (DFT).\cite{HohenbergPR64,KohnPR65}
While the properties calculated using the total energy (e.g., lattice constant)
are fairly well described by LDA or GGA, the electronic band
structure can not be described correctly in many cases.
For instance, it is well known that the
semilocal approximations (LDA and GGA) lead to unoccupied states in semiconductors
and insulators which are too low in energy, and hence, to too small band gap with
respect to experiment (see, e.g., Ref. \onlinecite{StadelePRL97}).
More advanced theories can provide more accurate band structures.
\cite{StadelePRL97,BechstedtPSSB09,AnisimovPRB91,BylanderPRB90,CoraSB04,TranPRL09}
The best known are the $GW$ approximation to the self energy $\Sigma_{\text{xc}}$
within the many-body perturbation theory (Ref. \onlinecite{HedinPR65})
and the hybrid functionals \cite{BeckeJCP93a,BeckeJCP93b} which consist of a
mixing of semilocal and Hartree-Fock (i.e., exact) exchange.

The $GW$ method represents the state-of-the-art,\cite{BechstedtPSSB09} and
leads very often to very accurate results in particular if it is applied self-consistently
\cite{FaleevPRL04,BrunevalPRL06,ShishkinPRL07} and with vertex corrections.
\cite{ShishkinPRL07} However, due to the huge computational effort required by
a $GW$ calculation, most of the $GW$ results from the literature were obtained
non-self-consistently (i.e., one-shot $G_{0}W_{0}$). Within the $G_{0}W_{0}$
method, the quasiparticle energies $\epsilon_{n\textbf{k}}^{G_{0}W_{0}}$ are
obtained as solutions of the \textit{nonlinear} equation
\begin{equation}
\epsilon_{n\textbf{k}}^{G_{0}W_{0}} =
\epsilon_{n\textbf{k}}^{\text{SL}} +
\langle\psi_{n\textbf{k}}^{\text{SL}}\vert
\Sigma_{\text{xc}}(\epsilon_{n\textbf{k}}^{G_{0}W_{0}}) -
v_{\text{xc}}^{\text{SL}}
\vert\psi_{n\textbf{k}}^{\text{SL}}\rangle,
\label{eGW}
\end{equation}
where $\psi_{n\textbf{k}}^{\text{SL}}$ and $\epsilon_{n\textbf{k}}^{\text{SL}}$
are the orbitals and the corresponding energies obtained from a previous
DFT calculation with a
semilocal (SL) functional $E_{\text{xc}}^{\text{SL}}$
[in Eq. (\ref{eGW}),
$v_{\text {xc}}^{\text{SL}}=\delta E_{\text{xc}}^{\text{SL}}/\delta\rho$].
The nondiagonal terms of the matrix of
$\Sigma_{\text{xc}}-v_{\text{xc}}^{\text{SL}}$ are neglected.\cite{HedinIJQC95}
The quasiparticle energies $\epsilon_{n\textbf{k}}^{G_{0}W_{0}}$ are
most of the time close to experiment, but there are known cases
(e.g., NiO\cite{vanSchilfgaardePRB06}) were self-consistency is really needed.
Beside the heavy cost of a $GW$ calculation ($G_{0}W_{0}$ can still be considered
as a very expensive method),
a drawback is that the convergence of the results with respect to
the number of unoccupied states can be extremely slow as, for example, for
ZnO.\cite{ShihPRL10,FriedrichPRB11}

The hybrid functionals are becoming more and more popular for solids (see, e.g.,
Ref. \onlinecite{CoraSB04}). On average, the accuracy of the electronic
band structure they provide is quite similar to $G_{0}W_{0}$.
In hybrid functionals, a fraction $\alpha_{\text{x}}$ of semilocal exchange
is replaced by the Hartree-Fock (HF) exchange:
\begin{equation}
E_{\text{xc}}^{\text{hybrid}} = E_{\text{xc}}^{\text{SL}} +
\alpha_{\text{x}}\left(E_{\text{x}}^{\text{HF}} -
E_{\text{x}}^{\text{SL}}\right).
\label{Exchybrid}
\end{equation}
The value of $\alpha_{\text{x}}$ which leads to the best agreement with
experiment depends on (a) the system under study, (b) the considered property,
and (c) the underlying semilocal functional $E_{\text{xc}}^{\text{SL}}$.
Most of the time, the value of $\alpha_{\text{x}}$ lies in the range 0$-$0.5.
Among the best known hybrid functionals, there is the so-called PBE0
\cite{ErnzerhofJCP99,AdamoJCP99} (the functional considered in the present work),
where the semilocal functional in Eq. (\ref{Exchybrid}) is the GGA of
Perdew, Burke, and Ernzerhof\cite{PerdewPRL96} (PBE) and
the amount $\alpha_{\text{x}}$ of Hartree-Fock exchange is set to 0.25
(Ref. \onlinecite{PerdewJCP96}). Another way of constructing hybrid
functionals consists of replacing only the short-range part of the semilocal
exchange by short-range Hartree-Fock as proposed in Ref. \onlinecite{HeydJCP03}
for the HSE functional (also based on PBE),
where the short- and long-range parts of exchange are defined by splitting
the Coulomb operator with the error function. Not considering the long-range
Hartree-Fock exchange avoids technical problems and makes the calculations
faster. Note that the idea of screening the Hartree-Fock exchange was already used
by Bylander and Kleinman for their screened-exchange LDA functional (sX-LDA),
\cite{BylanderPRB90} which can be considered as a hybrid functional with
100\% ($\alpha_{\text{x}}=1$) of
short-range Hartree-Fock exchange (see Ref. \onlinecite{ClarkPSSB11} for
recent sX-LDA calculations). However, even without long-range Hartree-Fock,
the use of hybrid functionals for solids leads to calculations
which are one or two orders of magnitude more expensive than with semilocal
functionals. The tendency of the PBE0 functional is to
overestimate small ($<3$ eV) band gaps and to underestimate large ($>10$ eV)
band gaps (see, e.g., Ref. \onlinecite{MarquesPRB11}),
and due to the neglect of the long-range Hartree-Fock in HSE, the HSE band gaps
are smaller than the PBE0 band gaps.\cite{HeydJCP05,PaierJCP06,MarquesPRB11}
Note that in Ref. \onlinecite{MarquesPRB11} it was shown that the performance of
PBE0 and HSE could be improved by making the fraction $\alpha_{\text{x}}$ of
Hartree-Fock exchange dependent on either the average of
$\left\vert\nabla\rho\right\vert/\rho$ in the unit cell
(as done in Ref. \onlinecite{TranPRL09} for the modified Becke-Johnson
potential\cite{BeckeJCP06}) or the static dielectric constant.

In this work, a fast way of getting the energies of orbitals from hybrid
functionals is proposed. In Ref. \onlinecite{TranPRB11}, the implementation
of hybrid functionals (screened and unscreened) into the WIEN2k code,
\cite{WIEN2k} which is based on the full-potential
linearized augmented plane-wave plus local orbitals method
\cite{AndersenPRB75,SjostedtSSC00} to solve the Kohn-Sham equations,
was reported. The Hartree-Fock method was implemented following the method
of Massidda, Posternak, and Baldereschi,\cite{MassiddaPRB93} which is based
on the pseudocharge method to solve the Poisson equation.\cite{WeinertJMP81}

The calculation of the matrix of the nonlocal Hartree-Fock operator
$\hat{v}_{\text{x}}^{\text{HF}}$ is done in a second variational procedure,
i.e., the operator $\alpha_{\text{x}}\left(\hat{v}_{\text{x}}^{\text{HF}}-
v_{\text{x}}^{\text{SL}}\right)$ is considered as a perturbation and the
semilocal orbitals are used as basis functions:
\begin{equation}
\langle\psi_{n\mathbf{k}}^{\text{SL}}\vert
\alpha_{\text{x}}\left(\hat{v}_{\text{x}}^{\text{HF}}-
v_{\text{x}}^{\text{SL}}\right)\vert
\psi_{n'\mathbf{k}}^{\text{SL}}\rangle.
\label{psivxHFvxSLpsi}
\end{equation}
The second variational procedure, which was also adopted for the
implementation of the Hartree-Fock equations in other LAPW codes
\cite{MassiddaPRB93,AsahiPRB99,BetzingerPRB10}
leads to cheaper calculations, since in practice
the number of orbitals $\psi_{n\mathbf{k}}^{\text{SL}}$ which are
used for the construction of Eq. (\ref{psivxHFvxSLpsi}) can be chosen to
be much smaller than the number of LAPW basis functions.
Since the shape of the orbitals obtained from the semilocal and hybrid functionals
can be expected to be rather similar, the most important matrix elements of
Eq. (\ref{psivxHFvxSLpsi}) are the diagonal ones. This leads to the simple
proposition which consists of calculating the orbital energies from hybrid
functional the following way:
\begin{equation}
\epsilon_{n\textbf{k}}^{\text{hybrid}} =
\epsilon_{n\textbf{k}}^{\text{SL}} +
\langle\psi_{n\mathbf{k}}^{\text{SL}}\vert
\alpha_{\text{x}}\left(\hat{v}_{\text{x}}^{\text{HF}}-
v_{\text{x}}^{\text{SL}}\right)\vert
\psi_{n\mathbf{k}}^{\text{SL}}\rangle,
\label{ehybrid}
\end{equation}
i.e., the nondiagonal terms are neglected, similarly as done in Eq. (\ref{eGW})
for the $G_{0}W_{0}$ method.\cite{HedinIJQC95}
Obviously, within this approximation the orbitals
are not updated since the matrix of the operator
$\alpha_{\text{x}}\left(\hat{v}_{\text{x}}^{\text{HF}}-v_{\text{x}}^{\text{SL}}\right)$
[Eq. (\ref{psivxHFvxSLpsi})] is diagonal, which means that the
$\psi_{n\mathbf{k}}^{\text{SL}}$s are already the eigenvectors.
In the following, the results obtained with Eq. (\ref{ehybrid}) will be named
PBE0$_{0}$ (i.e., one-shot PBE0) in analogy with $G_{0}W_{0}$. Compared to a
self-consistent calculation, using Eq. (\ref{ehybrid}) leads to
calculations of the band structure which are \textit{at least two orders of magnitude
faster} (neglect of the nondiagonal terms and only one iteration).
Naturally, the PBE orbitals were used as the semilocal orbitals in Eq. (\ref{ehybrid}).
Actually, Eq. (\ref{ehybrid}) is also very closely related to the way the exchange part
of the derivative discontinuity is calculated:
\begin{equation}
\Delta_{\text{x}} =
\langle\psi_{n'\mathbf{k}'}\vert
\hat{v}_{\text{x}}^{\text{HF}}-
v_{\text{x}}^{\text{EXX}}\vert
\psi_{n'\mathbf{k}'}\rangle -
\langle\psi_{n\mathbf{k}}\vert
\hat{v}_{\text{x}}^{\text{HF}}-
v_{\text{x}}^{\text{EXX}}\vert
\psi_{n\mathbf{k}}\rangle,
\label{deltax}
\end{equation}
where $\psi_{n\mathbf{k}}$ and $\psi_{n'\mathbf{k}'}$ are
the orbitals at the valence band maximum and conduction band minimum,
respectively, and $v_{\text{x}}^{\text{EXX}}$ is the multiplicative exact exchange
(EXX) potential obtained with the optimized effective potential method
[see Ref. \onlinecite{StadelePRL97} for more discussion on Eq. (\ref{deltax})].

In order to test the accuracy of Eq. (\ref{ehybrid}) for the calculation of
the orbital energies, we have considered the
following semiconductors and insulators (the structure and cubic lattice
constant are given in parenthesis): Ar (fcc, 5.260 \AA), C (diamond, 3.567 \AA),
Si (diamond, 5.430 \AA), GaAs (zinc blende, 5.648 \AA),
MgO (rocksalt, 4.207 \AA), NaCl (rocksalt, 5.595 \AA),
Cu$_{2}$O ($Pn\overline{3}m$, 4.27 \AA), MnO ($Fm\overline{3}m$, 4.445 \AA),
and NiO ($Fm\overline{3}m$, 4.171 \AA).
The unit cell of Cu$_{2}$O contains six atoms. Formally, Cu has a valency of $+1$ and
therefore the Cu-$3d$ shell is full, which means that the correlation effects
in the Cu-$3d$ shell should not play an important role as it is the
case for CuO.\cite{GhijsenPRB08}
MnO and NiO have a cubic symmetry, but by taking into account the antiferromagnetic phase
(along the [111] direction of the cubic cell), the symmetry is reduced to a
rhombohedral one (four atoms in the unit cell).
MnO and NiO are two of the most studied Mott insulators, for which the
semilocal functionals are very inaccurate due to the strongly localized character
of the $3d$ electrons.\cite{TerakuraPRB84}
Detailed descriptions of the structures of Cu$_{2}$O and MnO/NiO can be found in
Refs. \onlinecite{MarksteinerZPB86} and \onlinecite{CococcioniPRB05},
respectively.

\begin{table}
\caption{\label{table1}Transition energies (in eV) obtained with the PBE, PBE0,
and PBE0$_{0}$ methods. The experimental
value for Cu$_{2}$O is from Ref. \onlinecite{BaumeisterPR61}. See Table I of
Ref. \onlinecite{BetzingerPRB10} for the other solids.}
\begin{tabular}{llcccc}
\hline
\hline
Solid & Transition & PBE & PBE0 & PBE0$_{0}$ & Expt. \\
\hline
Ar        & $\Gamma\rightarrow\Gamma$     & 8.69 & 11.09 & 11.11 & 14.2        \\
C         & $\Gamma\rightarrow\Gamma$     & 5.59 &  7.69 &  7.64 &  7.3        \\
          & $\Gamma\rightarrow X$         & 4.76 &  6.64 &  6.59 &             \\
          & $\Gamma\rightarrow L$         & 8.46 & 10.76 & 10.73 &             \\
Si        & $\Gamma\rightarrow\Gamma$     & 2.56 &  3.95 &  3.90 &  3.4        \\
          & $\Gamma\rightarrow X$         & 0.71 &  1.91 &  1.87 &             \\
          & $\Gamma\rightarrow L$         & 1.53 &  2.86 &  2.80 &  2.4        \\
GaAs      & $\Gamma\rightarrow\Gamma$     & 0.53 &  1.99 &  1.87 &  1.63       \\
          & $\Gamma\rightarrow X$         & 1.46 &  2.66 &  2.65 &  2.18, 2.01 \\
          & $\Gamma\rightarrow L$         & 1.01 &  2.35 &  2.29 &  1.84, 1.85 \\
MgO       & $\Gamma\rightarrow\Gamma$     & 4.79 &  7.23 &  7.23 &  7.7        \\
          & $\Gamma\rightarrow X$         & 9.16 & 11.58 & 11.65 &             \\
          & $\Gamma\rightarrow L$         & 7.95 & 10.43 & 10.48 &             \\
NaCl      & $\Gamma\rightarrow\Gamma$     & 5.22 &  7.29 &  7.28 &  8.5        \\
          & $\Gamma\rightarrow X$         & 7.59 &  9.80 &  9.80 &             \\
          & $\Gamma\rightarrow L$         & 7.33 &  9.40 &  9.40 &             \\
Cu$_{2}$O & $\Gamma\rightarrow\Gamma$     & 0.53 &  2.77 &  2.68 &  2.17       \\
MnO       & $\Gamma\rightarrow\Gamma$     & 1.47 &  4.23 &  4.04 &             \\
NiO       & $\Gamma\rightarrow\Gamma$     & 2.41 &  6.07 &  5.91 &             \\
\hline
\hline
\end{tabular}
\end{table}

\begin{figure}
\includegraphics[scale=0.65]{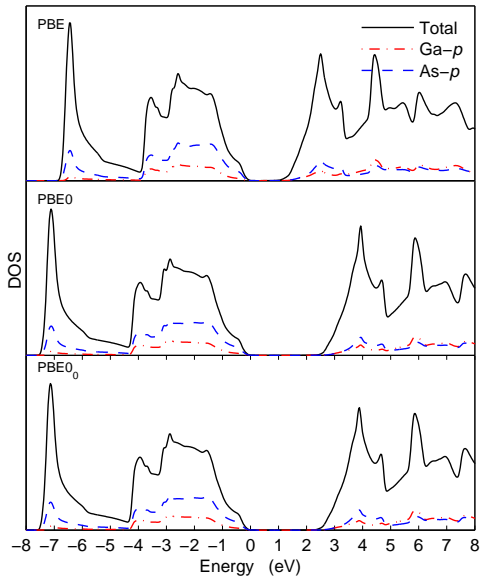}
\caption{\label{fig1}(Color online) Density of states of GaAs calculated with the
PBE, PBE0, and PBE0$_{0}$ methods. The Fermi energy is set at zero.}
\end{figure}
\begin{figure}
\includegraphics[scale=0.65]{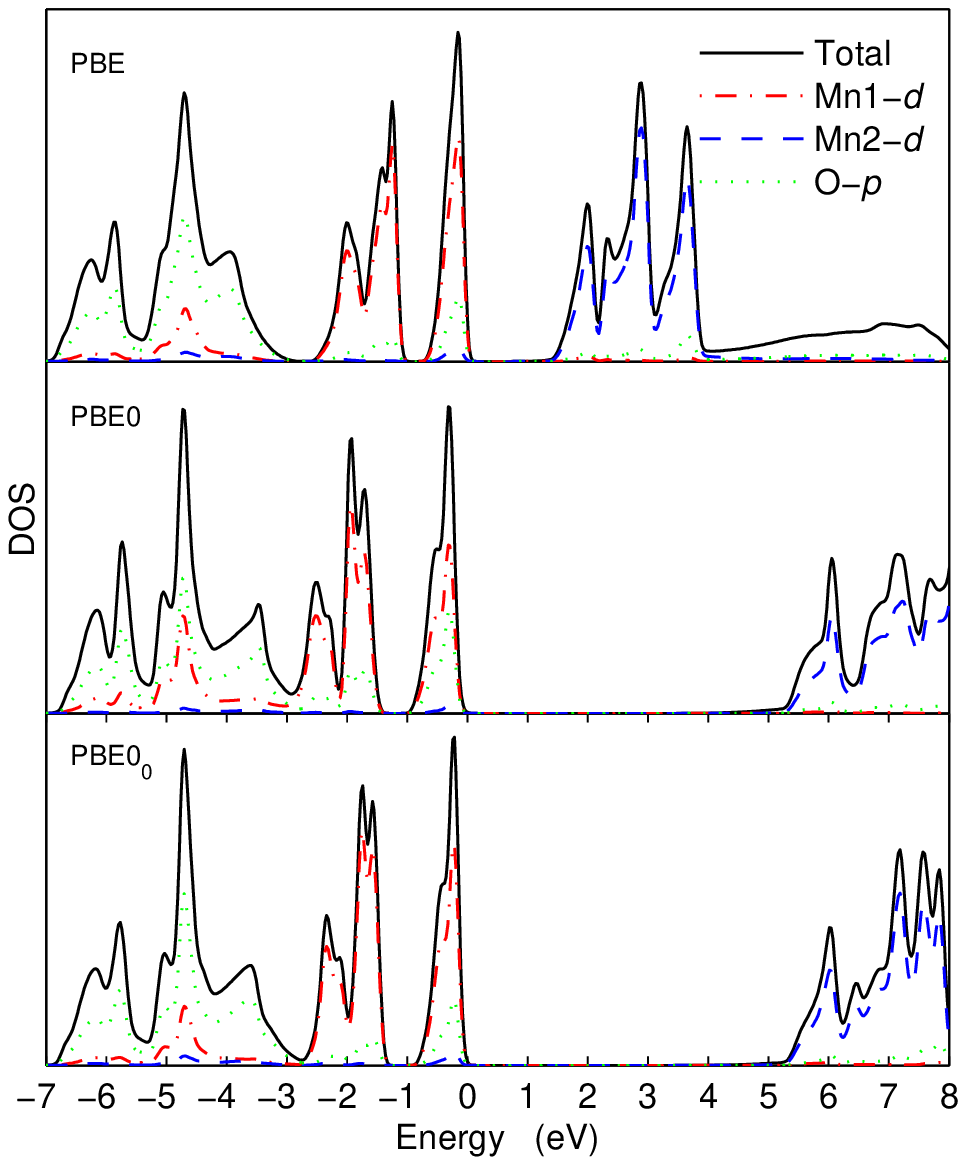}
\caption{\label{fig2}(Color online) Density of states of one spin component of
MnO calculated with the
PBE, PBE0, and PBE0$_{0}$ methods. The Fermi energy is set at zero.}
\end{figure}
\begin{figure}
\includegraphics[scale=0.65]{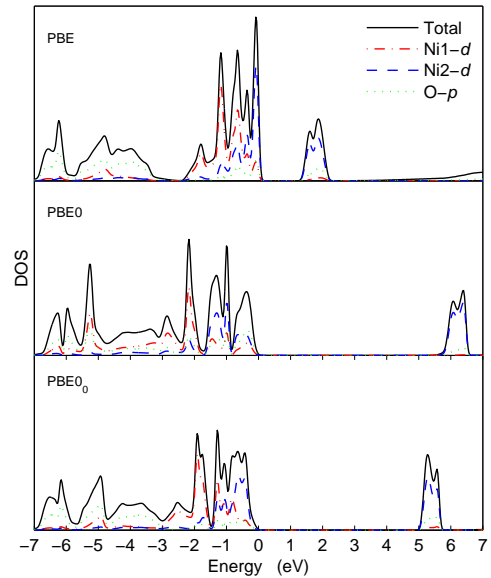}
\caption{\label{fig3}(Color online) Density of states of one spin component of
NiO calculated with the
PBE, PBE0, and PBE0$_{0}$ methods. The Fermi energy is set at zero.}
\end{figure}

The calculated transition energies, along with experimental values,
are shown in Table \ref{table1}. As expected, the PBE transition energies are
by far too small compared to the experiment, while the PBE0 values are much larger
and closer to the experiment. However, as already mentioned above, PBE0
overestimates small transition energies (e.g., GaAs) and underestimates large
transition energies (e.g., Ar). Actually, the important observation of the present work
is that the PBE0 transition energies calculated with the approximation
given by Eq. (\ref{ehybrid}) (PBE0$_{0}$ in Table \ref{table1})
are very similar to the self-consistent PBE0 results.
Indeed, the largest difference is for MnO ($\sim 0.2$ eV), which, anyway,
represents less than 10\% of the difference between PBE (1.47 eV) and PBE0 (4.23 eV).
For the $\Gamma\rightarrow\Gamma$ transition in GaAs and Cu$_{2}$O, the difference
between PBE0 and PBE0$_{0}$ is $\sim 0.1$ eV and less than 0.1 eV in all other cases.
In the case of MgO and NaCl, the accuracy of Eq. (\ref{ehybrid}) is impressive.

Figures \ref{fig1}, \ref{fig2}, and \ref{fig3} show the density of states (DOS)
of GaAs, MnO, and NiO, respectively. In the case of GaAs,
the PBE0 and PBE0$_{0}$ DOSs are indistinguishable, which can be explained by
the fact that already the PBE and PBE0 DOSs
are very similar and differ only by the shift of the unoccupied bands.
For MnO as well, the agreement between PBE0 and PBE0$_{0}$ is very good, albeit
small differences can be seen. In the case of NiO (the most difficult case
considered in this work), more visible differences between PBE0 and PBE0$_{0}$
can be observed. For instance, the PBE0$_{0}$ DOS in the energy range $-7$ and $-3$
eV looks intermediate between the PBE and PBE0 DOSs. There is less O-$p$ states in
the latter case. Also, the unoccupied Ni-$d$ peaks are higher in energy
by $\sim 0.8$ eV with PBE0 than with PBE0$_{0}$.

In Refs. \onlinecite{vanSchilfgaardePRB06} and \onlinecite{FaleevPRL04},
band gaps of 1.1 and 4.8 eV for NiO calculated with non-self-consistent
[Eq. (\ref{eGW})] and
self-consistent $GW$ calculations, respectively, were reported. The importance
of self-consistency in this case is rather extreme. Such a huge difference
is not observed for NiO with hybrid functionals when using
Eq. (\ref{ehybrid}) instead of doing a self-consistent PBE0 calculation.
The explanation lies in the fact that in the case of Eq. (\ref{ehybrid}),
only the semilocal orbitals are used for the construction of the nonlocal
Hartree-Fock operator $\hat{v}_{\text{x}}^{\text{HF}}$, while for
$G_{0}W_{0}$ [Eq. (\ref{eGW})] both the orbitals and their energies are used
for the calculation of the self-energy $\Sigma_{\text{xc}}$.
In the case of Cu$_{2}$O it was shown in Ref. \onlinecite{BrunevalPRL06},
that with respect to non-self-consistent $G_{0}W_{0}$, updating only the
orbital energies in $\Sigma_{\text{xc}}$ already increases the band by
0.46 eV (from 1.34 to 1.80 eV), while updating also the orbitals
increases further the band gap, but by a much smaller value (0.17 eV).
Other such examples can be found in Ref. \onlinecite{ShishkinPRB07}.
However, the effect of self-consistency in the $GW$ method can be reduced
if the input orbitals and energies are calculated from the
hybrid (see Ref. \onlinecite{BechstedtPSSB09}) or
$\text{LDA}+U$ (see Ref. \onlinecite{JiangPRB10}) methods.

In summary, it has been shown that for calculations with hybrid functionals,
neglecting the nondiagonal terms in the construction of the Hartree-Fock
Hamiltonian (within the second variational procedure)
leads to a very good approximation. By doing so, the orbital
energies [Eq. (\ref{ehybrid})] are very close to the energies obtained from a
self-consistent calculation done without any approximations.
Several types of semiconductors and insulators, including
the more difficult cases of antiferromagnetic MnO and NiO, have been considered and
the accuracy of the approximation has been shown to be (very) good in most cases.
For NiO, the results are less impressive than in the other cases.
The approximation leads to calculations of the band structure of solids with
hybrid functionals which are at least two orders of magnitude faster than the
self-consistent calculations. This approximation allows the calculation
of the electronic band structure of solids with hybrid functionals on much larger
systems.

\begin{acknowledgments}

Helpful discussions with Peter Blaha are greatly acknowledged.
This work was supported by Project SFB-F41 (ViCoM) of the Austrian Science Fund.

\end{acknowledgments}

\end{document}